\documentclass[%
 reprint,
superscriptaddress,
nofootinbib,
 amsmath,amssymb,
 aps,
]{revtex4-1}

\usepackage{graphicx}
\usepackage{dcolumn}
\usepackage{bm}
\usepackage{todonotes}

\newcommand{\rhat}{\hat{\mathbf{r}}}
\newcommand{\rvec}{\vec{\mathbf{r}}}
\newcommand{\K}{\mathcal{K}}
\newcommand{\GW}{\mathrm{GW}}

\begin{document}

\title{Cross-correlation of the astrophysical gravitational-wave background with galaxy clustering}

\author{Guadalupe Ca\~nas-Herrera}
\email{canasherrera@lorentz.leidenuniv.nl}
\affiliation{%
Leiden Observatory, Leiden University, PO Box 9506, Leiden 2300 RA, The Netherlands}
\affiliation{%
Lorentz Institute for Theoretical Physics, Leiden University, PO Box 9506, Leiden 2300 RA, The Netherlands
}%
\author{Omar Contigiani}
\email{contigiani@lorentz.leidenuniv.nl}
\affiliation{%
Leiden Observatory, Leiden University, PO Box 9506, Leiden 2300 RA, The Netherlands}
\affiliation{%
Lorentz Institute for Theoretical Physics, Leiden University, PO Box 9506, Leiden 2300 RA, The Netherlands
}%
\author{Valeri Vardanyan}
\email{valeri.vardanyan@ipmu.jp}
\affiliation{%
Leiden Observatory, Leiden University, PO Box 9506, Leiden 2300 RA, The Netherlands}
\affiliation{%
Lorentz Institute for Theoretical Physics, Leiden University, PO Box 9506, Leiden 2300 RA, The Netherlands
}
\affiliation{Kavli Institute for the Physics and Mathematics of the Universe (WPI), UTIAS, The University of Tokyo, Chiba 277-8583, Japan}

\date{\today}

\begin{abstract}
We investigate the correlation between the distribution of galaxies and the predicted gravitational-wave background of astrophysical origin. We show that the large angular scale anisotropies of this background are dominated by nearby non-linear structure, which depends on the notoriously hard to model galaxy power spectrum at small scales. In contrast, we report that the cross-correlation of this signal with galaxy catalogues depends only on linear scales and can be used to constrain the average contribution to the gravitational-wave background as a function of time. Using mock data based on a simplified model, we explore the effects of galaxy bias, angular resolution and the matter abundance on these constraints. Our results suggest that, when combined with galaxy surveys, the gravitational-wave background can be a powerful probe for both gravitational-wave merger physics and cosmology.
\end{abstract}

\maketitle


\section{\label{sec:intro}Introduction}

Gravitational waves (GWs) are one of the striking predictions of the General Theory of Relativity \cite{1916SPAW.......688E, 1918SPAW.......154E}. The first indirect detection was obtained by measuring the orbital decay of a pulsar binary system by Hulse and Taylor \cite{Hulse:1974eb} and, a century after they were conjectured, the GW signal of a merging black hole binary was detected by the Laser Interferometer Gravitational-Wave Observatory (LIGO) \cite{2016PhRvL.116f1102A}. Because the strain of GWs is less affected by distance compared to electromagnetic radiation, they potentially contain important information about sources which would be otherwise too dim to be observable. This discovery paved the way for a new multi-messenger era in cosmology and opened a new window into the physics of compact objects and gravity \cite{2018FrASS...5...44E}.

Every GW signal observed so far has been emitted from bright sources resolved as distinct events, such as low-redshift black hole \cite{2016PhRvL.116x1103A, 2017PhRvL.118v1101A, 2017ApJ...851L..35A, PhysRevLett.119.141101} and neutron star binary mergers \cite{2017PhRvL.119p1101A}. However, in addition to resolved events, one can expect the presence of a GW background (GWB) produced by the superposition of unresolved compact binaries that are either too far away or too faint to be detected individually. In practical terms, these unresolved sources form stochastic GWBs, which may differ in spectral shape and frequency depending on the source population \cite{2009PhRvD..80l2002T}.

For instance, massive black hole binaries form a stochastic background in the nHz band, which is expected to soon be detected by the Pulsar Timing Array (PTA) \cite{2016ApJ...819L...6T, Arzoumanian:2018saf, 2019ApJ...887...35M}. Because the mergers of this binary population are confined to the mHz band, the Laser Interferometer Space Antenna (LISA) is also expected to detect them as resolved events \cite{2003ApJ...590..691W}.

GWBs might also have a cosmological origin. Examples of such backgrounds are those produced in the early Universe, such as during inflation \cite{Starobinsky:1979ty}, or a phase transitions \cite{Blanco-Pillado:2017rnf}. Moreover, a hypothesized primordial black hole population \cite{Sasaki:2018dmp} might also contribute to the total number of compact binaries in the Universe. Many of these cosmological backgrounds are predicted to be isotropic and they can extend over multiple frequency bands, from nHz up to GHz \cite{2020arXiv200201079S, 2020PhRvD.101f3019W}. 

In this paper, we discuss the background due to solar-mass sized stellar remnants (black hole or neutron star binaries). The astrophysical GWB resulting from their inspiral and coalescence should be detectable not only in mHz band \cite{2019ApJ...871...97C}, but also in the Hz to kHz band. In this range, LIGO searches of this background have already been performed \cite{2019PhRvD.100f2001A}. 

While the experimental challenges associated with the detection of this GWB are not the focus of this work, it is worth pointing out that fundamental obstacles persist in both frequency ranges. In the mHz band, the reconstruction is hindered by the presence of an additional low-frequency background induced by Galactic white dwarf binaries \cite{2012CQGra..29l4016A}. To address this complication, previous works have shown that this background can be removed by exploiting the yearly modulation of space-based GW observatories \cite{2014PhRvD..89b2001A}. On the other hand, the main obstacle in the Hz-kHz is represented by the large shot noise contribution. Because the astrophysical GWB in this band is comprised of multiple unresolved transient events, a low event rate induces a large theoretical uncertainty in the total expected energy density. In particular, the contribution of this effect to the scale-dependence of the signal has a divergent formal expression \cite{Jenkins:2019uzp,Cusin:2018ump}.

None of these GWBs have been detected yet. Still, if ever observed, they would be the direct analogues of electromagnetic backgrounds formed by the superposition of multiple astronomical signals. Examples of this type of backgrounds are the cosmic infrared background (CIB) \cite{doi:10.1146/annurev.astro.39.1.249}, produced by stellar dust, and the cosmic X-ray background (CXB) \cite{doi:10.1146/annurev.aa.30.090192.002241}, formed by numerous extragalactic X-ray sources.

The anisotropies of the astrophysical GWB have been extensively studied for years  \cite{2009PhRvD..80l2002T} and, more recently, two independent groups \textit{Cusin et. al.} \cite{2018PhRvL.120w1101C, 2017PhRvD..96j3019C} and \textit{Jenkins et. al.} \cite{Jenkins:2018lvb, Jenkins:2018uac} obtained discrepant predictions for the scale-dependent signal \cite{Cusin:2018ump, Jenkins:2019cau}. The main disagreements are related to the shape of the angular power spectrum as well as the overall amplitude of the signal. The difference in shape seems to be related to the treatment of non-linear scales (see also Section \ref{sec:AGWB} of this paper), whereas the difference in amplitude is due to the chosen normalization. Here, let us mention that the main focus of their investigations so far has been the study of the autocorrelation signal and its shot-noise component, with \textnormal{further studies in this field being carried out also in~\cite{Cusin:2017mjm, Cusin:2019jhg, Pitrou:2019rjz, Cusin:2018avf}}. \textnormal{It is, however, worth pointing out that signals beyond autocorrelation, such as the cross-correlation between GWB and galaxy clustering or weak lensing convergence, have also been modelled to some extent (see \textit{e.g.} \cite{Cusin:2019jpv}).}

Here, we study the cross-correlation between the anisotropies of the astrophysical GWB and galaxy clustering (GC), and argue why it represents the ideal observable to detect the background and measure its properties. There are three main reasons for this choice. First, the distribution of compact mergers forming the GWB is determined by the distribution of their host galaxies. This means that one should expect a relatively large correlation between the two signals. Second, the cross-correlation signal for diffuse backgrounds is expected to have a larger signal-to-noise ratio compared to the autocorrelation signal, hence the former is likely to be detected earlier \cite{2014PhRvD..90b3514A}. Third, as presented in the next section, our investigation shows that the autocorrelation signal of the astrophysical GWB is very sensitive to small-scale structure, while the cross-correlation signal is free from this problem. In a somewhat similar spirit, Refs.~\cite{Mukherjee:2019wcg, Mukherjee:2019wfw, 2020arXiv200202466C} have recently studied the cross-correlation of \textit{resolved} GW sources with large scale structure and lensed cosmic microwave background.

Our paper is organized as follows. In Section~\ref{sec:AGWB} we review the main aspects of the GWB autocorrelation signal and highlight its limitations. In Section~\ref{sec:CC}, we present the angular power spectrum of the cross-correlation signal and calculate the expected shot-noise contamination (Appendix~\ref{app:noise}). In Section~\ref{sec:constraints} we demonstrate how the cross-correlation can be used to constrain the average power emitted by unresolved GW sources as a function of redshift, and quantify the required signal-to-noise ratio and angular resolution. To do this, we use a fiducial cosmology based on the best-fit results of Planck 2018 \cite{Aghanim:2018eyx} and a toy model for the GWB. Finally, we present our conclusions in Section~\ref{sec:conclusions}.

\section{gravitational-wave anisotropies}
\label{sec:AGWB}
In this section, we discuss the autocorrelation signal of the anisotropic GWB. This signal, as well as the shot-noise contamination, have been extensively studied in previous works \cite{Jenkins:2019uzp, Jenkins:2019nks, Cusin:2019jpv}. Here, we review the main aspects of modelling these and describe some particularities. 

Our starting point is the definition of the dimensionless energy density of GWs from a given direction of the sky $\rhat$, per unit solid angle:
\begin{equation}
    \Omega_{\GW}(\nu_0, \rhat) = \frac{\nu_0}{\rho_\mathrm{c}}\frac{\mathrm{d}\rho_{\GW}(\nu_0, \rhat)}{\mathrm{d}\nu_0 \mathrm{d}^2\rhat},
\end{equation}
where $\rho_{\GW}(\nu_0, \rhat)$ is the present-day energy density in GWs, $\nu_0$ is the observed frequency and $\rho_c=3H_0^2/8\pi G$ is the critical density of the Universe. Note that, from now on, we suppress the frequency dependence. We model this signal as 
\begin{equation}
    \Omega_{\GW}(\rhat) \equiv \int dr\; r^2 \K(r) n(\rvec),
    \label{eq:GWmap}
\end{equation}
where $n(\rvec)$ is the galaxy density field in comoving coordinates $\rvec $, and $\K$ is the GW kernel that encodes the average contribution of a galaxy to $\Omega_{GW}$ as a function of comoving distance $r$. In practice, this includes information about the star formation history of the Universe and the properties of the emitting binary population. It is instructive to rewrite Eq.~\eqref{eq:GWmap} in terms of the galaxy overdensity $\delta_\mathrm{g}(\rvec) \equiv n(\rvec)/\bar{n}(r) - 1$, with $\bar{n}(r)$ being the average number density of galaxies, defined as $\bar{n}(r) \equiv \int \mathrm{d}^2\rhat n(\rvec)/4\pi$. With this notation we have 
\begin{equation}
    \Omega_{\GW}(\rhat) = \int dr \; r^2 \K(r) \bar{n}(r) \left(\delta_\mathrm{g}(\rvec) + 1\right).
\end{equation}
From this point, the angular power spectrum of the anisotropic GWB $C_\ell^{\mathrm{GW}}$ can be calculated to be 
\begin{equation}\label{eq:autocorrel}
    C_\ell^{\mathrm{GW}} = 4\pi \int_{k_\mathrm{min}}^{k_\mathrm{max}} \frac{dk}{k} \vert\delta\Omega_\ell\vert^2 \mathcal{P}(k) + B_\ell^{\mathrm{GW}}.
\end{equation}
Here $\delta \Omega_\ell (k)$ is given by
\begin{equation}\label{eq:delta_Omega_ell}
    \delta \Omega_\ell (k) = \int dr  \; r^2\K(r)\bar{n}(r) T_g(k, r) j_\ell \left( kr\right),
\end{equation}
where $T_g$ is the synchronous gauge transfer function relating the galaxy power spectrum to the primordial one $P(k) = A_s \left(k/k_\ast\right)^{n_\mathrm{s} - 1}$, and $j_\ell$ is the spherical Bessel function of order $\ell$. Note that the galaxy bias is implicitly absorbed in $T_g$ and that we are neglecting relativistic corrections to $C_\ell^{\mathrm{GW}}$.  Note also that in Eq.~(\ref{eq:delta_Omega_ell}) we neglect relativistic corrections, as they are generally found to be below cosmic variance \cite{2019arXiv190911627B}. 

The term $B_\ell^{\mathrm{GW}}$ in the power spectrum is the shot-noise bias term introduced by the spatial and temporal shot-noise in the distribution of the individual events forming the GWB. Following \cite{Jenkins:2019uzp}, we write the shot-noise contribution in the kHz band as
\begin{equation}
    B_\ell^{\mathrm{GW}} = \int dr \; \K^2 (r) \bar{n}(r) r^2 \left[ 1+ \frac{1+z(r)}{R(r)T_O} \right].
    \label{eq:shotnoiseauto}
\end{equation}
Because of the low event rate in this frequency range, this noise contribution is inversely proportional to the average number of events per galaxy, written as the average redshifted event rate $(1+z)/R(r)$ multiplied by the observing time $T_O$. However, because the duration of the inspiral phase in the mHz band is much larger than any reasonable observing time, the contribution of the term $1/(R(r)T_0)$ is negligible in this case. 

The GWB discussed here is an integrated signal. Because of this, the low-redshift objects might significantly contribute to the GWB. Indeed, the astrophysical models of \cite{Cusin:2019jpv} suggest that the combination 
\begin{equation}\label{eq:Ktilde}
\tilde{\K}(r) =\K(r) \bar{n}(r)r^2 
\end{equation}
is not decaying to negligible values close to redshifts $z\sim0$. This introduces two complications in the modelling.

\begin{equation}
    B_\ell^{\mathrm{GW}} = \int dr \; \frac{\tilde{\K}^2(r)}{\bar{n}(r)r^2} \left[ 1+ \frac{1+z(r)}{R(r)T_O} \right].
\end{equation}

From this expression, it is clear that the shot-noise has a divergent expression due to low-redshift (low-$r$) contributions. To obtain a well-behaved prediction for the autocorrelation signal, this divergence can be suppressed if local events are excluded from the background. This is equivalent to setting a lower limit in the integral above different from zero.

Second, there exist a complication derived from the scale dependent part of the angular power spectrum (the first term in Eq.~\eqref{eq:autocorrel}), which is expected to receive non-negligible contributions from small, highly non-linear scales. To get some intuition about this feature, let us simplify our expression for the GWB angular power spectrum by using the so-called Limber approximation
\begin{equation}\label{eq:Limber}
    j_\ell(x) \rightarrow \sqrt{\frac{\pi}{2\alpha}}\delta_\mathrm{D}\left(\alpha - x\right),
\end{equation}
where $\delta_\mathrm{D}$ is the Dirac delta-function and $\alpha \equiv \ell + 1/2$. Using this in Eq. (\ref{eq:delta_Omega_ell}) and neglecting the bias term we obtain
\begin{equation}\label{eq:approx_autocorrel}
    C_\ell^{\mathrm{GW}} \approx \frac{2\pi^2}{\alpha}\int_{k_\mathrm{min}}^{k_\mathrm{max}}\frac{\mathrm{d}k}{k^3}\tilde{\mathcal{K}}^2\left(\frac{\alpha}{k}\right)\mathcal{S}^2\left(k,\frac{\alpha}{k} \right),
\end{equation}
\begin{equation}
\mathcal{S}(k, r) \equiv T_g(k, r) \mathcal{P}(k)^{1/2}.
\end{equation}

What Eq.~(\ref{eq:approx_autocorrel}) demonstrates is that $\tilde{\mathcal{K}}(r)$ acts as a modified kernel and selects a particular domain in the $k$-integral. This causes small scales to contribute significantly to $C_\ell^{\mathrm{GW}}$, unless $\tilde{\mathcal{K}}$ is vanishing at the lower end of its argument or $\tilde{\mathcal{S}}^2/k^3$ is falling fast enough at large values of $k$. As the modelling of the galaxy power spectrum at non-linear scales is highly uncertain, this feature is signalling a potential danger of using the autocorrelation signal as a probe of GW merger history or cosmology.    

\begin{figure*}[htp!]
    \centering
    \includegraphics[width=\columnwidth]{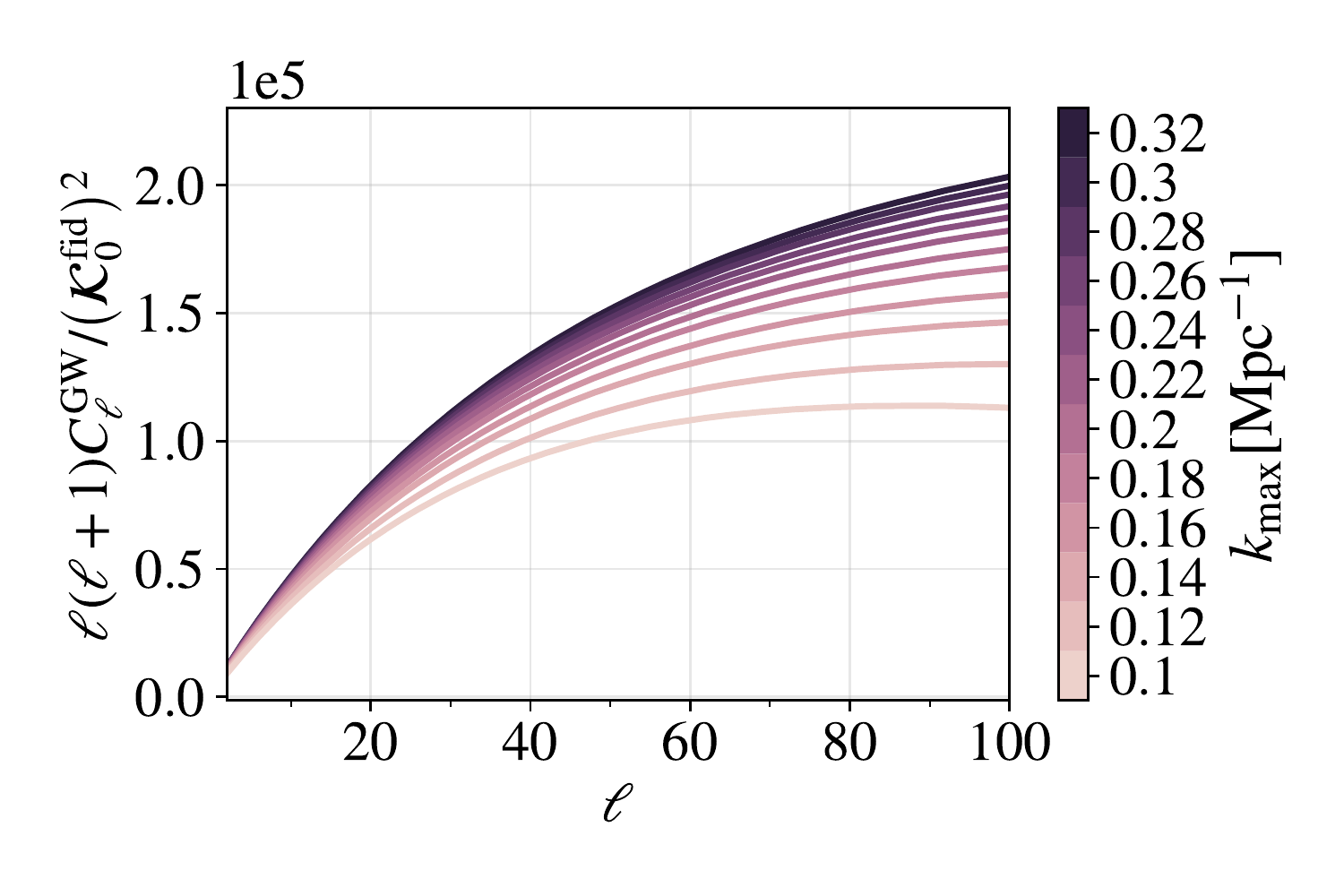}
    \includegraphics[width=\columnwidth]{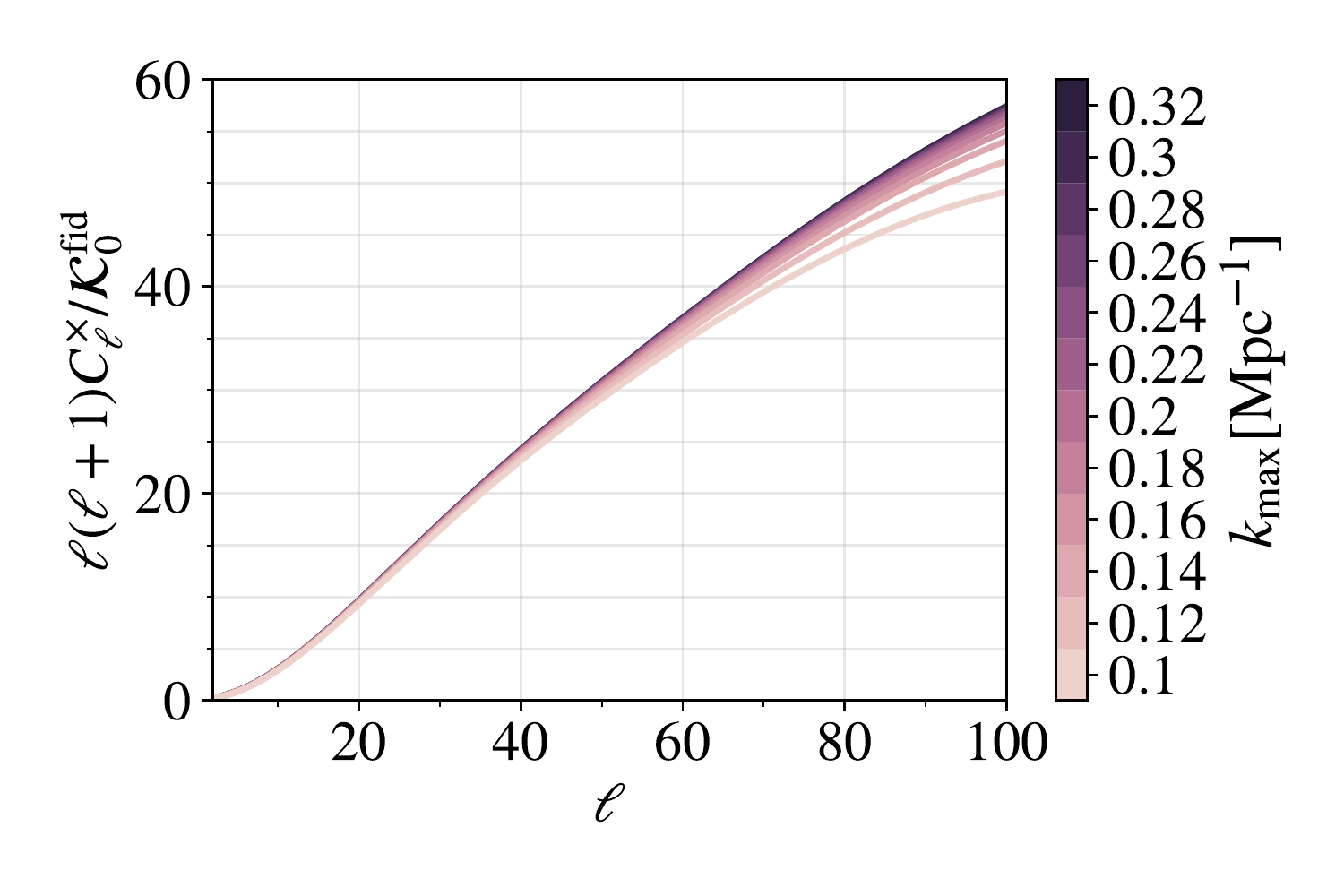}
    \caption{\textit{\textbf{Left panel:}} Linear autocorrelation power spectra $C_\ell^{GW}$ of the GWB of a constant $\tilde{K}(r)$ for a set of upper limits of the integral in Eq.~(\ref{eq:autocorrel}), in units of $\textrm{Mpc}^{-1}$.
    \textit{\textbf{Right panel:}} The same as in the left panel, but for the cross-correlation between a galaxy sample (centered at $z=0.5$) and the GWB, $C_\ell^{\times}$. Both of the panels are supposed to be understood as normalized with respect to the amplitude of the fiducial GWB model, to be discussed in detail later.}
    \label{fig:k-max-issue}
\end{figure*}

\vspace{10pt}

To accurately assess the impact of the issue mentioned above, let us turn to the results of exact numerical computations which do not rely on the Limber approximation. Having in mind the speed requirements of our later parameter analysis, we have developed a fast numerical procedure\footnote{The codes used in this paper are publicly available at \url{https://github.com/valerivardanyan/GW-GC-CrossCorr}.} to compute the integrals in Eqs.~(\ref{eq:autocorrel}) and (\ref{eq:delta_Omega_ell}), given the dark matter transfer function $T_\mathrm{m}(k,r)$ calculated using an Einstein-Boltzmann solver \footnote{In this paper, we use the $\Lambda$CDM limit of the \texttt{EFTCAMB} code \cite{Hu:2013twa, Raveri:2014cka} for simplicity, as it is easier to output the required transfer functions as a function of redshift.}.

A technical remark is in order here. Given the rapidly-oscillatory nature of the spherical Bessel functions in Eq.~(\ref{eq:delta_Omega_ell}), we have precomputed the line-of-sight integrals over these Bessel functions on bins of a fine $r$-grid. On the speed grounds, the source terms are then inserted only on a much coarser grid, which is only justified if these source functions do not vary significantly between two coarse-grid points. While this assumption is well justified for the transfer functions, we can only use our integrator if the kernel $\mathcal{K}(r)$ does not have rapid changes. In this paper, we consider only such smooth-enough kernels (and window functions -- see the next sections). We have verified the reliability of our integration procedure against a modified version of the latest public version of \texttt{CAMB} \cite{Lewis:1999bs, 2012JCAP...04..027H}. 

Our results are illustrated in the left panel of Fig.~\ref{fig:k-max-issue}, where we have chosen several values of $k_\mathrm{max}$, the upper limit of the integral in Eq.~(\ref{eq:autocorrel}), and calculated the corresponding angular power spectra for the multipoles in the range $\ell = [2, 100]$. Note in particular that the magnitude of the signal changes drastically with $k_\mathrm{max}$, meaning that the autocorrelation signal depends heavily on the shape of the low redshift power spectrum on non-linear scales. This is likely one of the causes behind the discrepancy between \textit{Jenkins et. al.} and \textit{Cusin et. al.} and suggests that an accurate prediction of the autocorrelation signal should take into account not only the shot-noise contribution \cite{Jenkins:2019nks,Cusin:2019jpv}, but also the uncertainties due to baryonic effects in the matter distribution at small scales \cite{2019arXiv190805765D, 2019JCAP...03..020S}. We point out, in particular, that the galaxy catalogue based on dark-matter-only simulations of \cite{2005MNRAS.360..159B} and the \texttt{halofit} model of \cite{2012ApJ...761..152T} are not designed to consistently or accurately model this uncertainty. While not shown, we point out that this problem is even more noticeable at high $\ell$, where a larger value of $k_\mathrm{max}\sim5$ Mpc$^{-1}$ is required for the integrals to converge (as highlighted in \cite{Cusin:2018ump}).

\section{cross-correlation with galaxy clustering}
\label{sec:CC}
In this section, we introduce the main concepts necessary for modelling the cross-correlation signal and discuss its advantages. 

First of all, we define the observed overdensity of galaxies in the given direction $\rhat$ per unit sold angle as 
\begin{equation}\label{eq:Dmap}
    \Delta (\rhat) = \int dr \; W_i(r) \delta_g(\rvec),
\end{equation}
where $W_i(r)$ is the probability density function of the galaxies' comoving distances (also referred to as the \textit{GC window function}) and $\delta_g(\rvec)$ is the galaxy overdensity defined earlier. Using Eq.~(\ref{eq:Dmap}), the angular power spectrum of GC, $C_\ell^{\mathrm{GC}}$, can be shown to be 
\begin{equation}\label{eq:ClGC}
    C_\ell^{\mathrm{GC}} =  4 \pi \int \frac{dk}{k} \vert \Delta_\ell(k) \vert^2 \mathcal{P}(k) + \frac{1}{n_i},
\end{equation}
where $\Delta_\ell(k)$ is given by
\begin{equation}\label{eq:GCsource}
    \Delta_\ell(k) = \int dr \; W_i(r) T_i(k, r) j_\ell (kr).
\end{equation}
$T_i(k, r)$ is the transfer function for the galaxy overdensity in the selected redshift range $W_i(r)$, $j_\ell (kr)$ is the spherical Bessel function of order $\ell$ and $n_i$ is the average number of galaxies per steradian, also dependent on the specific redshift selection $W_i(r)$. This final quantity appears the in second term in Eq.~(\ref{eq:ClGC}) and dictates the size of the shot-noise component of the power-spectrum. 

Using Eqs.~(\ref{eq:delta_Omega_ell}) and (\ref{eq:GCsource}), one can derive the angular power spectrum of the cross-correlation  $C_\ell^{\times}$ of the GWB and the GC maps, given by Eq.~(\ref{eq:GWmap}) and (\ref{eq:Dmap}). This is
\begin{equation}\label{eq:Cross-Corr-Cl}
    C_\ell^{\times} = 4 \pi\int \frac{dk}{k} \; \delta \Omega_\ell^\ast(k) \Delta_\ell(k) \mathcal{P}(k) + B_\ell,
\end{equation}
where the shot-noise contribution $B_\ell$, derived in Appendix~\ref{app:noise}, can be shown to be
\begin{equation}\label{eq:cross_shot_noise}
    B_\ell = \int dr \; W_i(r) \K(r).
\end{equation}
With these expressions in mind, we can now discuss how the cross-correlation signal can be used to address the modelling challenges we have presented in the previous section.

To address the first one, we notice that, while the $1/r^2$ divergence is still present in the integral in Eq.~\eqref{eq:cross_shot_noise}, this integral is generally well behaved if the window function $W_i(r)$ decays fast enough at small redshifts. Notice that this is impossible to do in the equivalent expression for the autocorrelation in Eq.~\eqref{eq:shotnoiseauto}.

With respect to the second issue, we compare in Fig.~\ref{fig:k-max-issue} the effects of the small-scale power spectrum on both the auto and cross-correlation. To explain the different behaviour, we note that the equivalent of Eq.~(\ref{eq:approx_autocorrel}) for the cross-correlation is
\begin{align}\label{eq:approx_crosscorrel}
    C_\ell^{\times} \approx \frac{2\pi^2}{\alpha}
    \int_{k_\mathrm{min}}^{k_\mathrm{max}}
    \frac{\mathrm{d}k}{k^3} 
    W_i\left(\frac{\alpha}{k}\right)
    \tilde{\mathcal{K}}\left(\frac{\alpha}{k}\right)
    \mathcal{S}^2\left(k,\frac{\alpha}{k} \right).
\end{align}
Because GC surveys allow for redshift-selection of the sources, the GC window function $W_i(r)$ can be taken to be peaked at some non-zero redshift and quickly decaying for larger or smaller values of $r$. Eq.~(\ref{eq:approx_crosscorrel}) shows that this behaviour cuts off the contribution from very large and very small scales, as shown in the right panel of Fig.~\ref{fig:k-max-issue}.

\section{Information content}
\label{sec:constraints}

\begin{figure}
    \centering
    \includegraphics[width=\columnwidth]{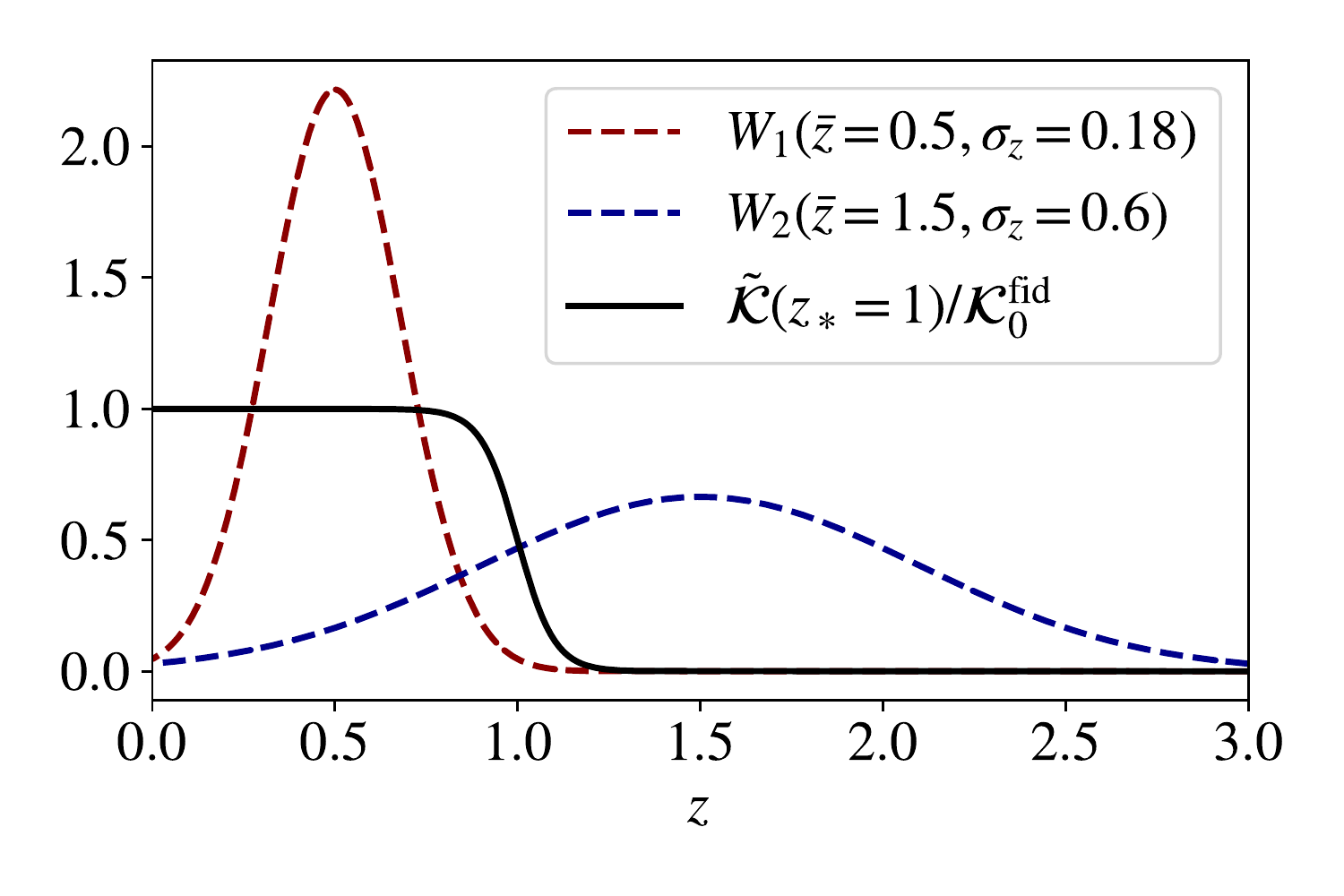}
    \caption{Fiducial model as a function of redshift $z$, of the GW source kernel $\tilde{\mathcal{K}}(r(z))$ in Eq. \eqref{eq:Ktilde}. In practice, we cut off the low-redshift sources with comoving distances smaller than $150\;\mathrm{Mpc}$ (see the text for details). The galaxy clustering window functions $W_1$ and $W_2$ are assumed to be Gaussian.}
    \label{fig:fiducial_model}
\end{figure}

\begin{figure*}[t!]
    \centering
    \includegraphics[width=\textwidth]{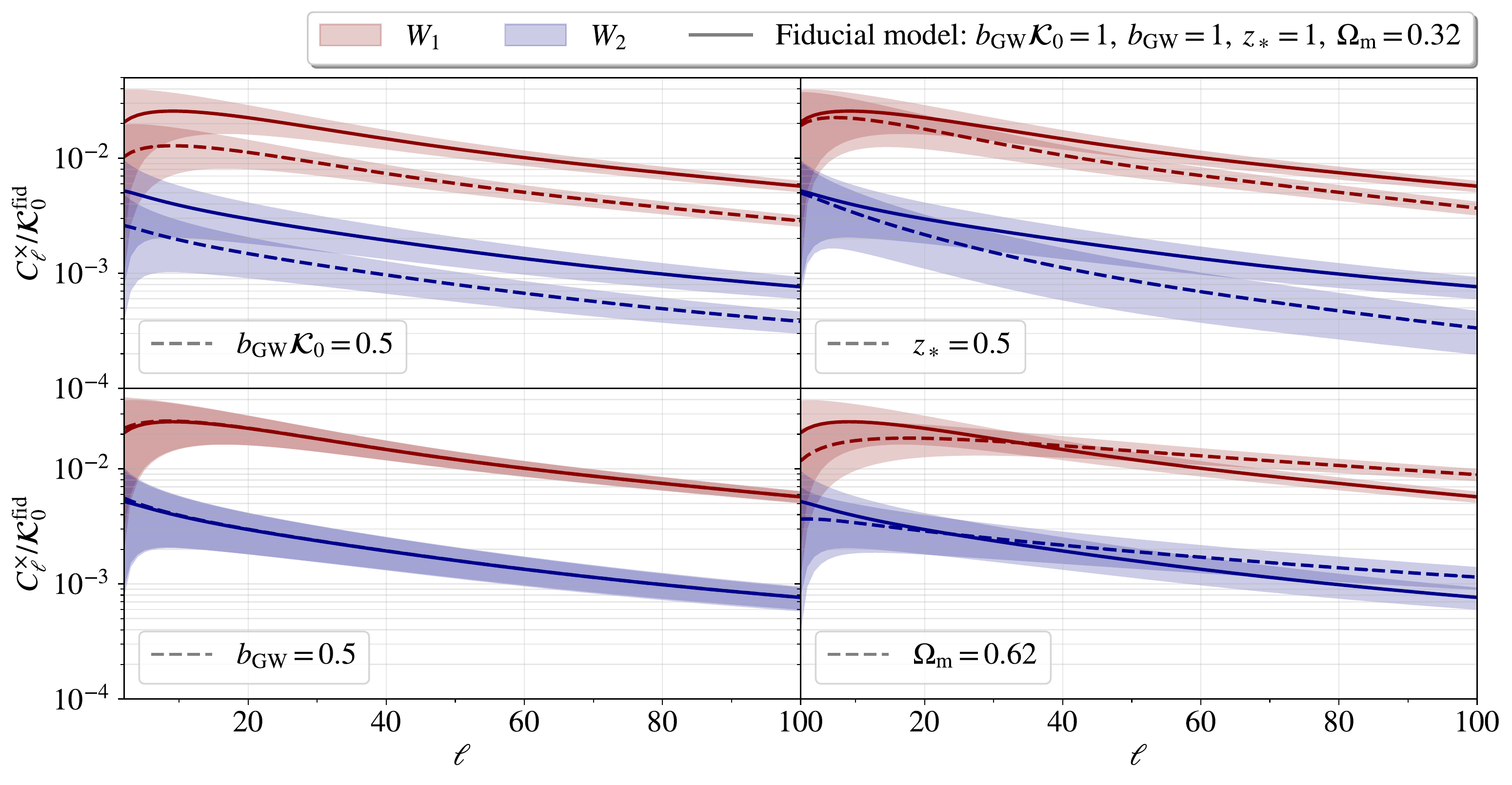}
    \caption{Effects of the model parameters $b_\mathrm{GW}\mathcal{K}_0, z_\ast, \Omega_\mathrm{m}$ and $b_\mathrm{GW}$ on the cross-correlation signal. The uncertainties are the cosmic variance defined in Appendix \ref{app:cosmic_variance}. Note particularly that in the case of both of the window functions $W_1$ and $W_2$ the change in $b_{\mathrm{GW}}\K_0$ induces a significant change in the amplitude of the signal (\textbf{\textit{upper left panel}}), while when the combination $b_\mathrm{GW}\mathcal{K}_0$ is fixed, the signal is not sensitive to the value of the GW bias $b_\mathrm{GW}$ (\textbf{\textit{lower left panel}}). Note that mostly the high-$\ell$ multipoles are sensitive to changes in $z_\ast$ (\textbf{\textit{upper right panel}}). Note also that the change on $\Omega_\mathrm{m}$ modifies the tilt of the signal, without altering its overall amplitude (\textbf{\textit{lower right panel}}). All of the panels are supposed to be understood as normalized with respect to the amplitude of the fiducial GWB model.}
    \label{fig:signal_response}
\end{figure*}

\begin{figure}[t!]
    \centering
    \includegraphics[width=\columnwidth]{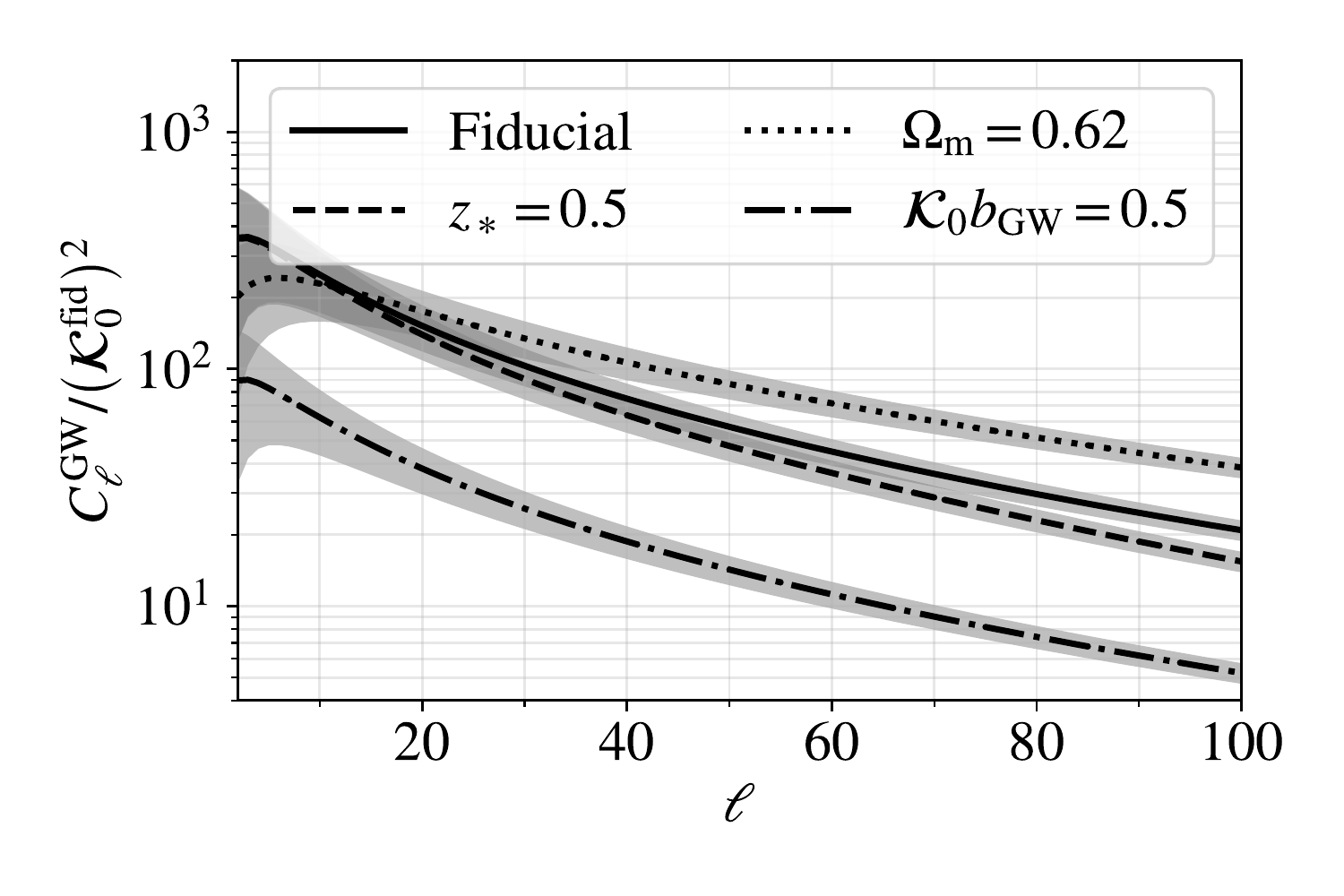}
    \caption{Effects of the model parameters $b_\mathrm{GW}\mathcal{K}_0, z_\ast, \Omega_\mathrm{m}$ and $b_\mathrm{GW}$ on the auto-correlation signal. The uncertainties are defined as in Fig.~\ref{fig:signal_response}.  The curves should be understood as normalized with respect to the amplitude of the fiducial GWB model.}
    \label{fig:autocorr_response}
\end{figure}

\subsection{Model set-up}\label{sec:model}
In this section, our primary goal is to explore the sensitivity of the cross-correlation signal to various parameters and estimate its information content. To this end, we model the signal using simple, but representative assumptions about the GW and GC maps. This allows us to derive an upper limit on the constraining power by assuming the theoretical minimum uncertainty due to cosmic variance.

We base our model for $\tilde{\K}(r)$ on the physically motivated one of \textit{Cusin et. al.} \cite{Cusin:2019jpv}, by noting that their function $\mathcal{A}(z)$ is the analogue of our $\tilde{\K}(r)$ in redshift space. In this reference, in particular, it is shown that $\mathcal{A}(r)$ is a slowly-evolving function of redshift, and has a similar shape over a wide range of frequencies and assumptions about the source population (see their figures 19 and 13). Thus, we model the kernel as
\begin{equation}\label{eq:K}
    \K(r) = \frac{\K_0}{2\bar{n}(r) r^2}\left\{ \tanh\left[10( z_\ast(r) - z(r))\right] + 1\right\},
\end{equation}
where $\K_0$ is the amplitude of the kernel, $z_\ast$ is a cut-off redshift, and $\bar{n} (r) \approx 10^{-1}$ $\mathrm{Mpc}^{-3}$ is the average comoving galaxy number density estimated using Figure~4 of \citep{Schaye:2014tpa}. We do not implement a redshift dependence for this quantity because its value is relevant only for the shot-noise component of the cross-correlation, found to be negligible in the cases considered here. In our fiducial model, we assume $z_\ast=1$ (see Fig. \ref{fig:fiducial_model}), as it known by \textit{Cusin et al.} that the astrophysical kernel $\K(r) \bar{n}(r)r^2$ is expected to decay around that value in redshift. Notice that, while $\mathcal{K}_0$ should be dimensionful, its units are irrelevant to us because the cross-correlation signal is proportional to its value. For the rest of the paper, we call $K_0^\text{fid}$ the fiducial value of this quantity.

In the next subsections, we study the cross-correlation between the GWB modelled above and two galaxy catalogues centred at different redshifts. The two window functions, $W_1$ and $W_2$, are assumed to be Gaussian distributions centered at $\bar{z}=\{0.5, 1.5\}$ and with widths of $\sigma_z = \{0.18, 0.6\}$. These values are picked so that the two selections overlap with the constant portions of $\tilde{\K}(r)$.

Moreover, we model the transfer functions in Eqs.~\eqref{eq:GCsource} and \eqref{eq:delta_Omega_ell} by using a linear bias approximation (valid for large scales):
\begin{equation}
T_i(k) = b_i T_\mathrm{m}(k, r),
\end{equation}
and
\begin{equation}
T_g(k,r)= b_\mathrm{GW}T_\mathrm{m}(k, r),
\end{equation}
where $T_\mathrm{m}(k, r)$ is the transfer function for cold dark matter and the $b_X$ are known as bias parameters. When varying our model, we freeze the bias of both galaxy catalogues since it can be extracted from the clustering autocorrelation signal alone. On the contrary, we treat the GW bias $b_\mathrm{GW}$ as a free parameter and we assume it to be a constant over redshift. While this is not necessarily true, in the absence of shot-noise, only the combination $b_\mathrm{GW}\tilde{\K}(r)$ appears in the signal. This implies that a more complex model can always capture any redshift dependence through the function $\tilde{\K}(r)$. Note, however, that breaking the degeneracy between the linear bias of the GW population and the amplitude of the astrophysical kernel $\K(r)$ requires a full understanding of the GWB kernel and all ingredients \cite{Scelfo_2018}.\\

For the rest of the analysis, we focus on the mHz frequency band, and assume that low-redshift events (below $r=150$ Mpc) can be filtered. In our modelling, as discussed in the previous sections, these assumptions are essential to obtain a well-behaved signal which is not overwhelmed by noise. For reference, under these assumptions we get the following relative noise values at $\hat{\ell}=10$:
\begin{equation}
    \frac{B^\mathrm{GW}_{\hat{\ell}}}{C_{\hat{\ell}}^\mathrm{GW}}\approx
    \frac{B_{\hat{\ell}}}{C^{\times}_{\hat{\ell}}} \approx 10^{-4}.
\end{equation}
The first value is derived using the inspiral time of a solar mass black hole binary starting from $1$ mHz \cite{1995PhRvL..74.3515B}, an observing time of $1$ year and a merger rate of $10^{-5}$ per year \cite{LIGOScientific:2018mvr}. 

As a summary of our model, Fig.~\ref{fig:fiducial_model} contains the two window functions $W_1, W_2$ and the kernel $\tilde{\K}(r)$.

\subsection{Behaviour of the cross-correlation}

Before attempting to reconstruct the parameters of our model from mock data, let us gain some insights into the response of the cross-correlation signal on various parameters.

First, we explore the dependence of the signal on the kernel amplitude $\K_0$, or, more precisely, the combination $b_\mathrm{GW}\K_0$. In the upper left panel of Fig.~\ref{fig:signal_response} we can see that in the case of both of the window functions $W_1$ and $W_2$ the change of the amplitude induces a significant change in the signal. Note that here the bias itself is fixed. In reality, the kernel amplitude $\K_0$ and the bias are perfectly degenerate with each other since the two appear as proportionality constants to both cross-correlation signals. To see this, in the lower left panel of Fig.~\ref{fig:signal_response} we demonstrate the impact of varying $b_\mathrm{GW}$ on the signal when $b_\mathrm{GW}\K_0$ is held fixed. Note that a similar scaling with the kernel amplitude is present also for the autocorrelation signal shown in Fig.~\ref{fig:autocorr_response}, which is proportional to $(b_\mathrm{GW}\K_0)^2$.

Second, we turn our attention to the dependence of the signal on the turnover redshift $z_\ast$. In the upper right panel of Fig.~\ref{fig:signal_response} we see that the change of $z_\ast$ induces a change in the shape of the signal. The signal with $W_2$ is sensitive to $z_\ast$, while in the case of $W_1$ the signal is practically independent of it. A similarly small effect is also visible in the autocorrelation signal in Fig.~\ref{fig:autocorr_response}.

Third, it is interesting to show the effect of $\Omega_\mathrm{m}$ on the signal. Specifically, in the lower left panel of Fig.~\ref{fig:signal_response}, it is demonstrated that the effects of $\Omega_\mathrm{m}$ and $\K_0$ are qualitatively different from each other. Indeed, changing $\Omega_\mathrm{m}$ rotates the signal, while $\K_0$ affects the amplitude of the signal. This rotation effect due to varying $\Omega_\mathrm{m}$ is expected, as a similar effect is observed in the galaxy clustering autocorrelation signal. Indeed, such a behaviour in the signal allows galaxy clustering to constrain both $\Omega_\mathrm{m}$ and the normalization of the matter power spectrum $\sigma_8$ (see e.g. \cite{Blanchard:2019oqi}).

Finally, we point out that the scale-dependent power spectra discussed in this sections do not have a clear peak for any value of $\ell$ and practically do not show any sign of decaying power for small scales. This is in contrast to the naive expectations based on galaxy clustering result. This difference is due to the interplay between projected scales and redshift selection described in Section \ref{sec:AGWB}, together with the use of relatively wide effective window functions ($\tilde{K}(r)$, $W_1$ and $W_2$).

\subsection{Constraining $\K(r)$}

\begin{figure*}[hpt!]
    \centering
    \includegraphics[width=\columnwidth]{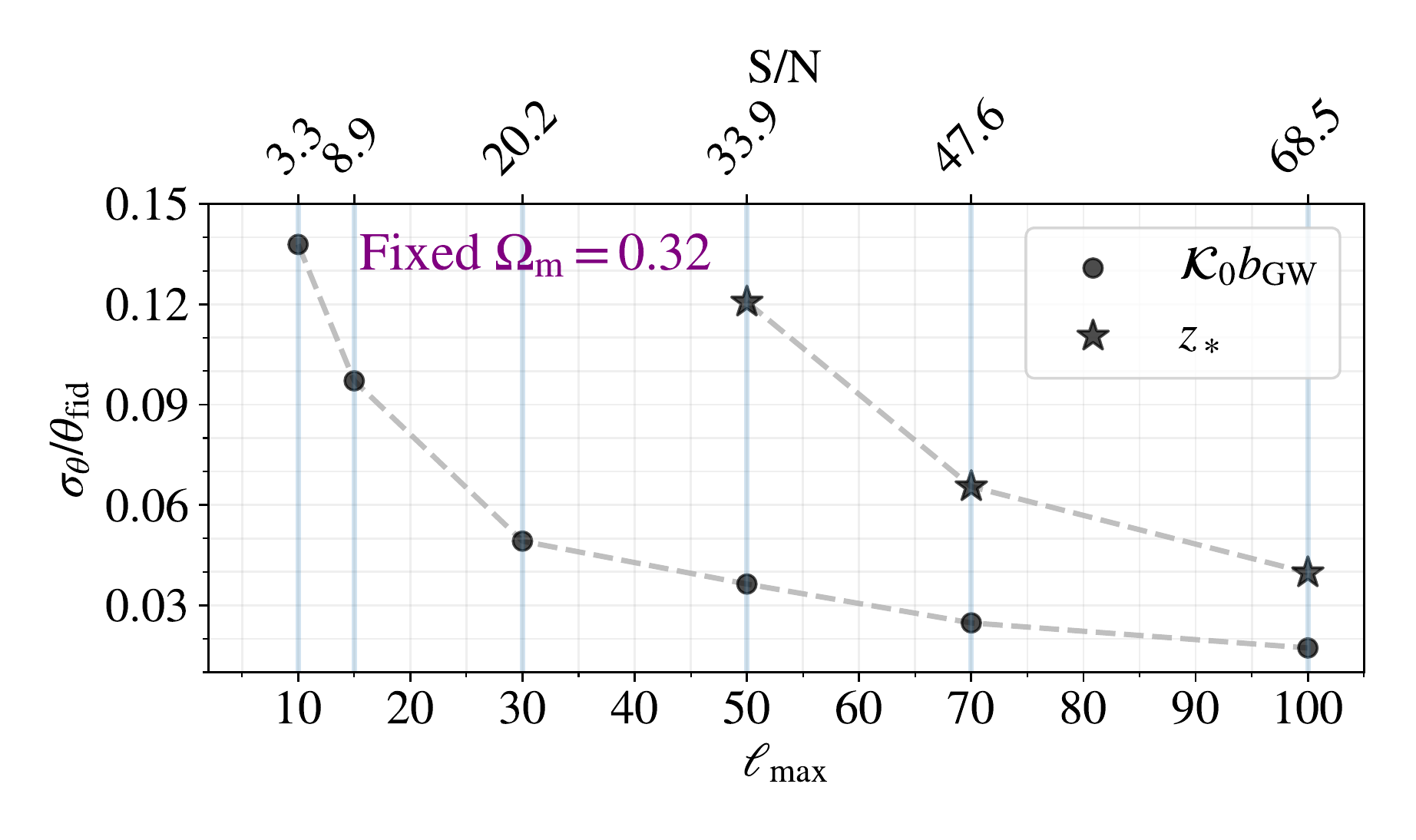}
    \includegraphics[width=\columnwidth]{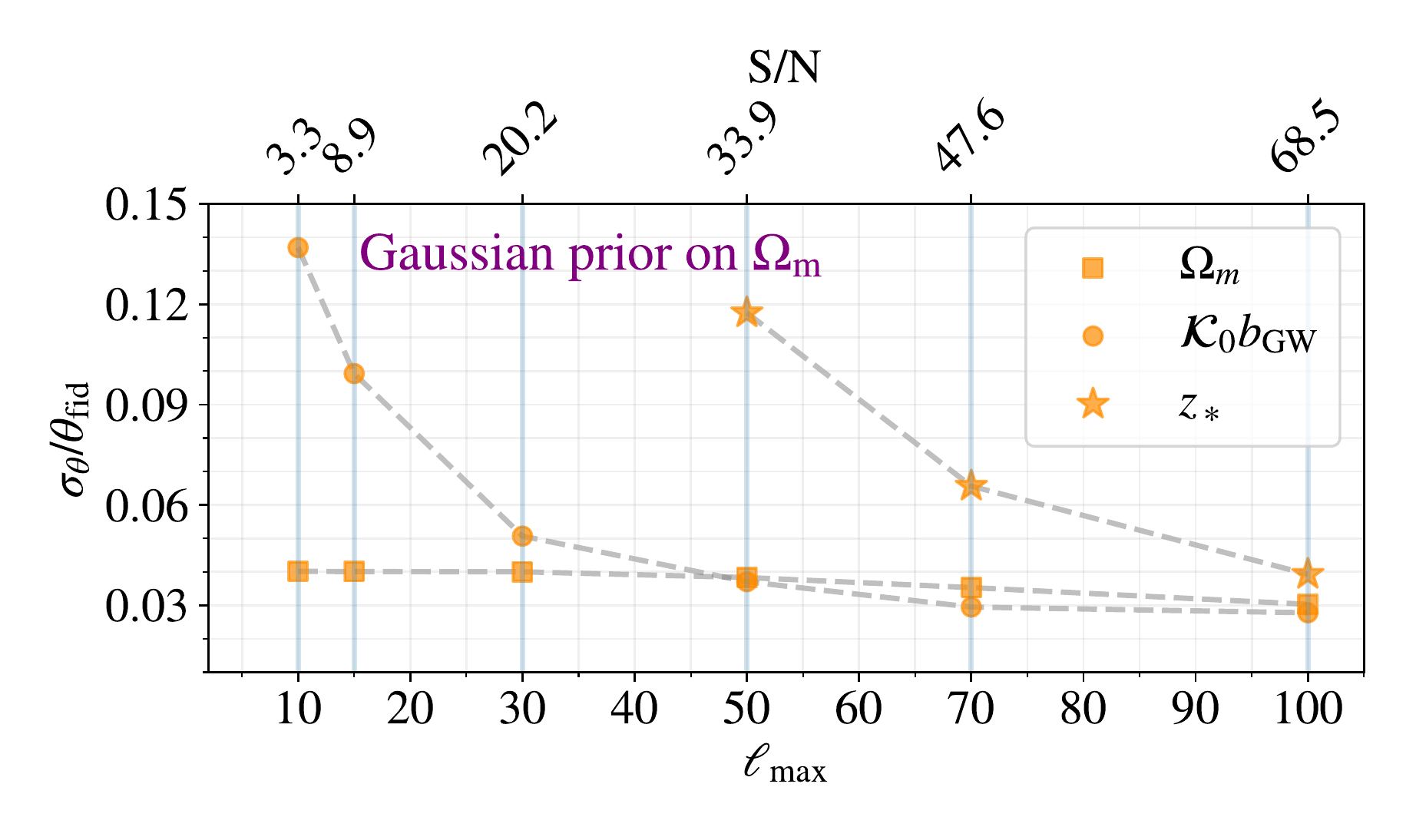}
    \caption{Constraints on the GWB parameters ($b_\text{GW}\K_0,z_\ast$) and cosmology ($\Omega_m$) obtained using the cross-correlation signal with two window functions as a function of the maximum multipole included in the analysis. Cosmic-variance limited measurements are assumed for all the constraints, so these should be understood as the best-case scenario results. Larger values of the signal-to-noise ratio (S/N) correspond to better angular resolution (see Eq.~\ref{eq:SN}). We have explored the effect of $\Omega_\mathrm{m}$ on these constraints by either fixing its value (\textbf{\textit{left panel}}), or setting a Planck-2018-like Gaussian prior (\textbf{\textit{right panel}}). Remarkably, the combination $b_\mathrm{GW}\mathcal{K}_0$ can be constrained even with very limited angular sensitivity. The turnover location $z_\ast$ is practically unconstrained for $\ell_\mathrm{max}\lesssim 50$, and $\Omega_\mathrm{m}$ is prior dominated for these multipoles. In case of $\ell_\mathrm{max}\gtrsim 50$ all the relevant parameters are tightly constrained, and for $\ell_\mathrm{max}\sim 100$ the constraints are at the level of a few percent. Notably, the cosmology (mimicked by varying $\Omega_\mathrm{m}$ in our analysis) can match and surpass the CMB results only in case of high angular resolution/signal-to-noise. For reference, $\ell_\text{max}=100$ roughly corresponds to $2$ degrees.}  
    \label{fig:results_long}
\end{figure*}

The goal of this section is to understand the constraining power of the cross-correlation signal by studying how precisely the astrophysical model can be inferred from a noisy $C_\ell$ measurement.

In our analysis, we focus on the best-case scenario of cosmic-variance limited uncertainties as derived in Appendix \ref{app:cosmic_variance}
and use a simple proxy for the overall signal-to-noise ratio of the cross-correlation, defined as
\begin{equation}\label{eq:SN}
    \left( \frac{\mathrm{S}}{\mathrm{N}}\right)^2 \equiv \sum^{\ell_\mathrm{max}}_{\ell = \ell_\mathrm{min}} \frac{ \left(C_\ell^\times\right)^2
    }{\mathrm{Var}
    C_\ell^\times }.
\end{equation}
In the presence of multiple, independent window functions, we simply sum the relative signal-to-noise expressions in quadrature. 

We compute the cross-correlation power spectra, given in Eq.~(\ref{eq:Cross-Corr-Cl}), using the model presented in Section~\ref{sec:model}, and attempt to recover the model parameters from a noisy realization. To explore the inferred constraints as a function of angular resolution and S/N levels, we do this in several multipole ranges of $\ell$ with $\ell_\text{min}=2$ and varying $\ell_\text{max}$.

The parameters of interest in our analysis are the amplitude of the GWB kernel $\K_0$ and the turnover redshift $z_\ast$. In addition to these, we also explore the bias $b_\mathrm{GW}$ and $\Omega_\mathrm{m}$ to see if variations in $T_g(k, r)$ can affect the inferred $\K(r)$, and to explore the possible degeneracies between the GWB model and cosmology. To include the effects of varying $\Omega_\mathrm{m}$ we have precomputed the dark matter transfer functions for a grid of $\Omega_\mathrm{m}$ values, and have inferred the results for the intermediate values through nearest neighbour interpolation. 

The exploration of the parameter space is carried out using the MCMC \texttt{python} code \texttt{emcee} \cite{emcee}. We have employed a Gaussian likelihood function on $C_\ell$ with diagonal covariance matrix given through Eq.~(\ref{eq:cov}), and the prior ranges given in Table~\ref{tab:prior}. Note that since we expect $\mathcal{K}_0$ to be degenerated with $b_\mathrm{GW}$, we do not vary $\mathcal{K}_0$ itself, but rather vary the combination $b_\mathrm{GW}\mathcal{K}_0$.   

\begin{table}[h!]
\begin{tabular}{c|c|c}
Parameter & Fiducial value & Prior \\
\hline
$b_\mathrm{GW}\K_0$ & 1 &  $[0.01, 100]$  \\
$b_\mathrm{GW}$ & 1 &  $[0.1, 10]$  \\
$z_\ast$ & 1 &  $[0.5, 1.5]$  \\
$\Omega_\mathrm{m}$ & 0.32& $\mathcal{G}(0.32, 0.013)$ \\
\hline
\end{tabular}
\label{tab:prior}
\end{table}

The main results of the analysis are summarised in Fig.~\ref{fig:results_long}, where we show the expected constraints on the parameters of interest as a function of the maximum multipole included in the analysis. We also show the corresponding cosmic-variance-only signal-to-noise ratios.

\begin{figure}[hpt!]
    \centering
    \includegraphics[width=1.\columnwidth]{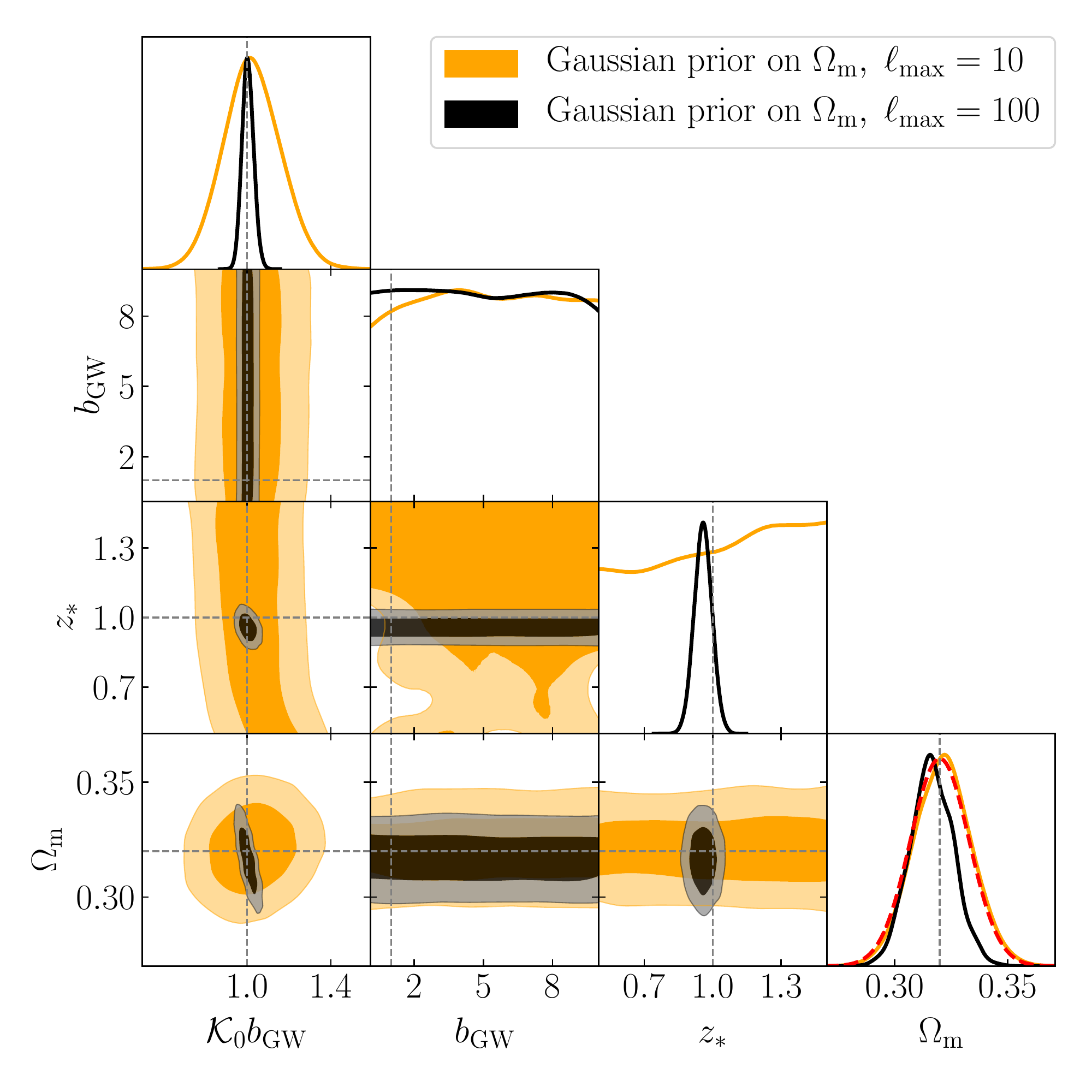}
    \caption{Posterior distributions for the cases of $\ell_\mathrm{max} = 10$ (orange) and $\ell_\mathrm{max} = 100$ (black), with Planck-2018-like Gaussian prior on $\Omega_\mathrm{m}$ (shown in red dashed line). Contours represent the $68\%$ and $95\%$ confidence regions. We can clearly see that $\K_0 b_\mathrm{GW}$ is constrained even in the case of the limited angular resolution, while $b_\mathrm{GW}$ is never separately constrained. The turn-over redshift $z_\ast$ is unconstrained for the low-resolution case, while it is tightly constrained for the case of $\ell_\mathrm{max} = 100$. Finally, $\Omega_\mathrm{m}$ is prior dominated for the low-resolution case, while it beats the prior in the high-resolution scenario. Also noteworthy are the degeneracies between $\Omega_\mathrm{m}$ and $\K_0 b_\mathrm{GW}$, as well as between $z_\ast$ and $\K_0 b_\mathrm{GW}$.}  
    \label{fig:results_corner}
\end{figure}

Let us first have a look at the left panel of the figure, which corresponds to a fixed $\Omega_\mathrm{m}$ value. As we see, $b_\text{GW}\K_0$ is constrained and, notably, this is true even in the limited multipole range corresponding to $\ell_\mathrm{max} = 10$. This is expected, as a clear detection of the signal is associated with a measurement of its amplitude.  On the other hand, less encouraging are the results for the turnover redshift $z_\ast$, which can be constrained only for $\ell_\text{max}\gtrsim 50$ or, equivalently, a S/N of $\sim 33$.


In the right panel of the figure, we now impose a Gaussian prior on $\Omega_\mathrm{m}$, with its variance being comparable to the Planck-2018 constraint on $\Omega_\mathrm{m}$. While the $z_\ast$ results are not affected, the uncertainties on the amplitude are now slightly inflated, due to a degeneracy between $\Omega_\mathrm{m}$ and $b_\text{GW}\K_0$. This is also visible in the signal responses plotted in Fig.~\ref{fig:signal_response}.

Let us now fully concentrate on the two limiting angular sensitivities in our analysis. We take a LIGO-like angular sensitivity limited to the multipole range of $\ell\in [2, 10]$, as well as an angular sensitivity of a hypothetical high-resolution GW detector corresponding to $\ell\in [2, 100]$. The full constraints, for the case of Gaussian priors on $\Omega_\mathrm{m}$, are presented in Fig.~\ref{fig:results_corner}. 

We can clearly see that $\K_0 b_\mathrm{GW}$ is constrained even in the case of the limited angular resolution, while $b_\mathrm{GW}$ is never separately constrained. We have checked that the latter feature is also present in all the other runs presented in this Section. This justifies our choice to vary the combination $b_\mathrm{GW}\mathcal{K}_0$ instead of varying $b_\mathrm{GW}$ and $\mathcal{K}_0$ separately. 

The turn-over redshift $z_\ast$ is unconstrained for the low-resolution case, while it is tightly constrained for the case of $\ell_\mathrm{max} = 100$. The dark matter abundance $\Omega_\mathrm{m}$ is prior dominated for the low-resolution case, while it beats the prior in the high-resolution scenario. Also noteworthy are the degeneracies between $\Omega_\mathrm{m}$ and $\K_0 b_\mathrm{GW}$, as well as between $z_\ast$ and $\K_0 b_\mathrm{GW}$. These can be easily understood by inspecting the combined behaviours presented in Fig.~\ref{fig:signal_response}.

Before turning to our conclusions let us mention that the results presented in this section depend on the precise details of the GC window functions and GWB detection and more precise results can only be obtained by performing a realistic forecast with exact survey/detector specifications. While we leave a more detailed investigation for future research, our results suggest that a cosmic-variance limited measurement of the GWB anisotropies down to $\ell\sim100$ is able to tightly constrain the redshift evolution of the GW kernel $\tilde{\K}$, as well as to provide Planck-like constraints on cosmological parameters. 

\section{Conclusions}
\label{sec:conclusions}
In this paper, we have discussed in detail the angular power spectrum of the cross-correlation between the GWB of astrophysical origin and GC.

We have shown that, contrary to the autocorrelation signal, the cross-correlation signal does not depend heavily on the small-scale galaxy power spectrum and hence is a more robust observational probe. To this point, we have also shown that the shot-noise associated with this signal is small for realistic choices of the window functions $W_i$.

Then, armed with these results, we studied in detail the properties of the angular power spectra for a range of model parameters. In particular, we have shown how the signal is sensitive to the turnover redshift $z_\ast$ of the GWB kernel, a combination of its amplitude and the bias $b_{\mathrm{GW}}\mathcal{K}_0$, as well as the dark matter abundance $\Omega_\mathrm{m}$. We have also shown that the signal is not separately sensitive to $b_{\mathrm{GW}}$ and $\mathcal{K}_0$. A summary of these is presented in Fig.~\ref{fig:signal_response}.

As one of the main goals of this paper, we have performed a Bayesian parameter estimation using an MCMC sampling based on mock data with cosmic-variance-limited uncertainties. This choice allows us to provide an upper limit on the constraining power of this new observational probe (Fig.~\ref{fig:results_long}). In particular, we have demonstrated that the cross-correlation signal is a powerful tool to constrain the properties of the GWB kernel $\K(r)$ if appropriate GC window functions are used. This is true even when marginalizing over uncertainties in the cosmology gravitational-wave bias.

We have quantified for the first time the need of high-resolution GW detectors in order to extract the full information content of the GWB of astrophysical origin. In particular, we have shown that both a high angular resolution and a high signal-to-noise ratio ($\ell \sim 100$, $\text{S/N} \sim 70$) are required to recover both the matter abundance $\Omega_\mathrm{m}$ and features of the kernel $\K(r)$ as a function of redshift. Note, in particular, that these requirements are far above the angular resolution of present-day and near-future detectors (roughly $\ell\lesssim 10$). While this is not the priority of currently proposed third-generation detectors \cite{Maggiore_2020}, it is worth noting that the advantages of high-resolution gravitational-wave astronomy are numerous and not limited to the study of this anisotropic background \cite{Baker:2019ync}.

The case for studying the cross-correlation is strengthened by noticing that the anisotropies of the GWB in kHz band will most probably first be measured through cross-correlation with galaxy surveys, as the latter will provide a guiding pattern to be looked at in the noisy GW data. Given the promising nature of our results regarding the constraints of the GW kernel parameters and $\Omega_\mathrm{m}$, we believe that the cross-correlation between GW and GC has the potential to be a robust observational probe in the era of multi-messenger cosmology.

\acknowledgements
It is our pleasure to thank Giulia Cusin, Alexander C. Jenkins, Eiichiro Komatsu, Matteo Martinelli and Mairi Sakellariadou for useful discussion. We also thank Ana Ach\'ucarro for useful comments and encouragement. OC and VV acknowledge financial support through de Sitter PhD fellowship of the Netherlands Organization for Scientific Research (NWO). GCH acknowledges support from the Delta Institute for Theoretical Physics (D-ITP consortium), a program of the Netherlands Organization for Scientific Research (NWO). The work of VV is also financed by WPI Research Center Initiative,
MEXT, Japan

\appendix 
\section{Shot-noise for the cross-correlation signal}
\label{app:noise}

We follow \cite{Jenkins:2019uzp} and evaluate the shot-noise contribution to the observed cross-correlation signal $C_\ell^{\times}$ in terms of the shot-noise contribution to the covariance between the observed maps $\Omega(\rhat)$ and $\Delta(\rhat^\prime)$. Our starting point is
\begin{equation}
    B_\ell = \int d^2 \rhat P_\ell(\rhat \cdot \rhat^\prime) \mathrm{Cov}[\Omega(\rhat), \Delta (\rhat^\prime)]_{\mathrm{SN}}.
    \label{eq:A1}
\end{equation}

By keeping in mind that $\tilde{\K}(r) = r^2 \K (r)\bar{n}(r)$ and that $\delta_g(\rvec) = \left(n(\rvec)-\bar{n}(r)\right)/\bar{n}$ we use the definitions in Eqs.~(\ref{eq:GWmap}),~(\ref{eq:Dmap}) to write:
\begin{align}
    &\mathrm{Cov}[\Omega(\rhat), \Delta (\rhat^\prime)]_{\mathrm{SN}}  = \nonumber\\ &\int dr \int dr^\prime \frac{r^2}{\bar{n}} \times 
    \mathrm{Cov}[\K(r)n(\rvec), W_i(r^\prime)n(\rvec^\prime)]_{\mathrm{SN}}.
    \label{eq:Cov}
\end{align}

As a side note, we point out that this expression is a stretch of notation since, formally, the quantities $\K(r)n(\rvec)$ and $W(r) n(\rvec)$ represent the mean values of the variables that we are trying to correlate. To proceed, we notice that $W(r) n(\rvec)$ is proportional to the number density of galaxies visible in the galaxy survey and that $\K(r)n(\rvec)$ is proportional to the number density of GW events around an infinitesimal volume centred in $\rvec$. This is confirmed by the formalism used in the aforementioned references \cite{Jenkins:2019uzp} and \cite{2018PhRvL.120w1101C} to predict a realistic $\K(r)$. 

In a finite volume $\delta V_i$ we write down the number of GW mergers as 
\begin{equation}
    \Lambda_i = \sum^{N_i}_k \lambda_k ,
\end{equation}
where $N$ is the number of galaxies present in this volume and the $\lambda_j$-s are the number of events for each galaxy. If we assume that $N$ and $\lambda_k$ are Poisson distributed, $\Lambda_i$ follows a compound Poisson distribution with variance
\begin{equation}
    \mathrm{Var}[\Lambda_i] = \langle \Lambda_i^2 \rangle - \langle \Lambda_i \rangle^2  = \langle N_i \rangle \left(\langle \lambda \rangle + \langle \lambda \rangle^2\right).
\end{equation}

If we call $f$ the fraction of galaxies in the volume $\delta V_j$ visible in the galaxy survey we also derive:

\begin{equation}
    \mathrm{Cov}[fN_j, \Lambda_i] = f \langle N \rangle\langle \lambda \rangle \delta_{ij},
\end{equation}
where $\delta_{ij}$ is the Kronecker delta. By going back to the continuous case, we obtain the following result:

\begin{equation}
    \mathrm{Cov}[\K(r)n(\rvec), W_i(r^\prime)n(\rvec^\prime)]_{\mathrm{SN}} = \bar{n}(r) W_i(r) \K(r) \delta^3(\rvec-\rvec^\prime).
\end{equation}

Finally, by plugging everything into Eq.~\eqref{eq:A1} we obtain the result shown in the main text:

\begin{equation}
    B_\ell = \int dr ~W_i(r) \K(r).
\end{equation}

\section{Cosmic variance of the cross-correlation signal}
\label{app:cosmic_variance}

Assume we have two maps on the sky, corresponding to the GWB and GC anisotropies. The angular decomposition coefficients $a^{\mathrm{GW}}_{\ell m}$ and $a^{\mathrm{GC}}_{\ell m}$ are assumed to be Gaussian random variables with zero mean, and each $m$-mode is drawn from the same distribution. The relevant angular power spectra are defined as

\begin{align}
    C^{\times}_\ell &\equiv \mathrm{Cov}\left[ a^{\mathrm{GW}}_{\ell m}, a^{\mathrm{GC}}_{\ell m}\right],\\
    C^{\mathrm{GW}}_\ell &\equiv \mathrm{Var}\left[ a^{\mathrm{GW}}_{\ell m}\right],\\
    C^{\mathrm{GC}}_\ell &\equiv \mathrm{Var}\left[ a^{\mathrm{GC}}_{\ell m}\right].
\end{align}

It is then trivial to construct an unbiased estimator of the cross-correlation power spectrum as
\begin{align}
    \widehat{{C^{\times}_\ell }} = \frac{1}{2\ell + 1} \sum_{m = -\ell}^{+\ell}a^{\mathrm{GW}}_{\ell m}a^{\mathrm{GC}}_{\ell m}.
\end{align}

The variance of this estimator can then be shown to be
\begin{align}
    \mathrm{Var} C^{\times}_\ell = \frac{1}{\left(2\ell + 1\right)^2} \sum_{m = -\ell}^{+\ell}\mathrm{Var}\left[a^{\mathrm{GW}}_{\ell m}a^{\mathrm{GC}}_{\ell m}\right] =\nonumber\\ 
    \frac{1}{\left(2\ell + 1\right)^2} \sum_{m = -\ell}^{+\ell} C^{\mathrm{GW}}_\ell C^{\mathrm{GC}}_\ell +\nonumber\\ \mathrm{Cov}\left[\left(a^{\mathrm{GW}}_{\ell m}\right)^2, \left(a^{\mathrm{GC}}_{\ell m}\right)^2\right] - \mathrm{Cov}\left[a^{\mathrm{GW}}_{\ell m}, a^{\mathrm{GC}}_{\ell m}\right]^2.
\end{align}

In summary, we have
\begin{align}\label{eq:cov}
    \mathrm{Var}C^{\times}_\ell = 
    \frac{
        C_\ell^\mathrm{GW} C_\ell^\mathrm{GC} + 
        \left(C_\ell^\times\right)^2
    }{2\ell + 1},
\end{align}
where we have used the Gaussianity of $a_{\ell m}$'s. Making the $a^{\mathrm{GC}}_{\ell m} \rightarrow a^{\mathrm{GW}}_{\ell m}$ replacement turns this expression into
\begin{align}
    \mathrm{Var}C^{\mathrm{GW}}_\ell = \frac{2
        \left(C^{\mathrm{GW}}_\ell \right)^2
    }{2\ell + 1},
\end{align}
which, of course, recovers the usual cosmic variance result. 

\bibliography{GWGC}

\end{document}